\documentclass[twocolumn,showpacs,preprintnumbers,amsmath,amssymb]{revtex4}
\usepackage{graphicx}
\usepackage{dcolumn}
\usepackage{bm}

\begin{document}
\title{CP nonconservation in the leptonic sector}
\author{Petre Di\c t\u a}
\altaffiliation{
 Institute of Physics and Nuclear Engineering, 
P.O. Box MG6, Bucharest, Romania}

\begin{abstract}
In this paper we use   an exact method to impose unitarity on moduli  of   neutrino PMNS matrix recently determined, and show how one could obtain information on CP nonconservation from a limited experimental information. One suggests a novel type of global fit by expressing all  theoretical quantities in terms of convention independent parameters: the Jarlskog invariant $J$ and the moduli $|U_{\alpha i}|$, able to resolve the positivity problem of $|U_{e 3}|$. In this way the fit will directly provide a  value for  $J$, and if it is different from zero it will prove the existence of  CP violation in the available experimental data.
If the best fit result, $|U_{e3}|^2<0$, from M. Maltoni {\em et al}, [New J.Phys. {\bf 6} (2004) 122] is confirmed, it will imply a new physics in the leptonic sector. 
\end{abstract}

\pacs{12.15.Hh,  12.15.-y, 14.60.Pq~~~~~~~~~~~~}DOI: 10.1103/PhysRevD.74.113010


\maketitle

\section{Introduction}

Neutrino experiments have shown that neutrinos have mass and oscillate, the last property suggesting a mixing of  leptons similar to that of quarks \cite{SNO}-\cite{K2K};
 for more references on experimental data see, e.g., \cite{MSTV}-\cite{FLMP}. The presence of the lepton mixing opens the possibility to see a CP violation in the leptonic sector, phenomenon which is considered  an indispensable ingredient that could generate the excess of the baryon number of Universe \cite{ADS}, since the baryon number can be generated by leptogenesis \cite{FY}.  Excluding the LSND anomaly, see  \cite{CA}-\cite{AA}, all  neutrino data are explained by three flavor neutrino oscillations, and the determination of oscillation parameters is one of the  main goals of  phenomenological analyzes. The LSDN signal for $\bar{\nu}_{\mu}\rightarrow \bar{\nu}_e$ has been interpreted as a possible exstence of a fourth light neutrino, or as a signal of $CPT$ violation.  The current opinion is that neither explanation provides a good fit to all existing neutrino data, see e.g. \cite{BMW}.
Recently two detailed combined analyzes of all neutrino data have been published, see Refs. [5-6],   that materialized in upper and lower bounds on $\sin\theta_{ij}$, these bounds being converted in Ref.\,\cite{GG} into intervals for the moduli of the PMNS unitary matrix \cite{MNS}. In both  above papers the CP phase $\delta$ was not considered  a free parameter, e.g., only the particular cases $\cos\delta=\pm 1$ have  been considered in \cite{FLMP}. We shall write the numerical moduli matrix as
\begin{eqnarray}
V=\left(\begin{array}{cll}
0.835\pm 0.045&0.54\pm 0.07&  < 0.2\\
0.355\pm 0.165& 0.575\pm 0.155&0.7\pm 0.12\\
0.365\pm 0.165&0.590\pm 0.15&0.685\pm 0.125\\
\end{array}\right),\,\, \label{mns1}
\end{eqnarray}
where the central values are given by half the sums of lower and upper bounds  entering \cite{GG}, and the errors correspond to a three sigma  level, excepting  $V_{e3}$ for which only an upper bound is given. Complementary information concerns the absolute value of the two neutrino mass-squared differences \cite{MSTV}, but we have no information on:
\vskip2mm
$\bullet$ the magnitude of  $U_{e3}$-element of the leptonic mixing matrix, or the value of the mixing angle $\theta_{13}$;
\vskip2mm
$\bullet$ the existence of leptonic  CP violation, or a value for the Dirac phase $\delta\ne0,\,\pi$ entering the PMNS matrix \cite{MNS};
\vskip2mm
$\bullet$ the order of  the mass spectrum, or the sign of  $\Delta m^2_{13}$;
\vskip2mm
$\bullet$ the magnitude of  neutrino masses, etc.; see, e.g., \cite{HMS}-\cite{OM}.
\vskip2mm 
Although it is believed that only the next generation experiments will give an answer to the first above items, \cite{HM}, we consider that by using  unitarity we can obtain some information from  experimental data as given in (\ref{mns1}) for both the existence of CP violation in  leptonic sector, and the parameter ~$\theta_{13}$, all the more as  measurement of $\theta_{13}$ is considered  a  first mandatory ingredient for  investigation of CP leptonic violation in  $\nu_{\mu}\rightarrow \nu_e$ transitions and for the mass hierarchy determination. Concerning  CP violating phase, $\delta$, the general opinion is that it will require a major experimental effort because of the intrinsic difficulty to measure it, see e.g. Ref.\,\cite{GMMT}.  

The current phenomenological analyzes of data are given in terms of mixing angles $\theta_{ij}$, see, e.g., \cite{MSTV}-\cite{FLMP}, and phase $\delta$ which all  depend  on the chosen parametrization of the unitary matrix one starts with. In the paper we propose another method for analyzing the data by expressing all measurable quantities directly related to oscillation amplitudes in terms of moduli, $|U_{\alpha i}|$, and Jarlskog invariant $J$, \cite{J}. From a theoretical point of view this approach has the advantage that the reconstruction of a unitary matrix from  knowledge of moduli leads essentially to a unique solution, see, e.g., Ref. \cite{D}.  From an experimental point of view the method allows to get from nowadays limited information on neutrino physics all  moduli of the unitary matrix such that one can predict the approximate form for the oscillation amplitudes, for example $\nu_{\tau}\rightarrow \nu_{\mu}$, which could be of great help for experimenters. It is easily seen that the experimenters have measured quantities 
 directly related to oscillation amplitudes, which contain information  on the moduli $ |U_{e1}|,\,\, |U_{e2}|,\,\, |U_{\mu3}|,\,\, {\rm and}\,\ |U_{\tau3}|$. The above moduli are correlated  by unitarity to $ |U_{e3}|$
\begin{eqnarray}|U_{e3}|^2=1-|U_{e1}|^2-|U_{e2}|^2=1-|U_{\mu3}|^2-|U_{\tau3}|^2 \label{ec} \end{eqnarray}
 By using (\ref{ec}), from the first row and from the third column  in  (\ref{mns1}) one gets the bounds
\begin{eqnarray}\begin{array}{c}-0.1465\le |U_{e3}|^2\le 0.155,\\*[2mm]
 -0.3285\le |U_{e3}|^2\le 0.35\end{array} \end{eqnarray} 
 The moduli of lower and upper bounds in the above relations are of the same order of magnitude, which means that the possibility $V_{e3}^2\le 0 $ is not  excluded.  In fact the best fit of the KamLAND data in Ref.\,\cite{MSTV} gave $V_{e3}^2=-0.13$;  see also Figure A.2 in Ref.\,\cite{MSTV} and, respectively  Figure 14 in Ref.\,\cite{FLMP}.  Hence the phenomenologists have to take into account this possibility, which, if confirmed, could be interpreted as a breaking of the established rules, signaling  really a new physics. A slight modification of the  theory can take it properly into account. After nonvanishing masses, such a surprise coming from neutrino physics is not {\em a priori} excluded.

As we will show in the paper our approach makes full use of  unitarity property, which implies many correlations between all moduli of a unitary matrix. For a complete determination of an (exact) unitary matrix  the knowledge of  four independent moduli is necessary, see, e.g., Ref.\,\cite{D}. Taking into account that a $3\times 3$ unitary matrix
has nine moduli, the choice of four ones is completely at our disposal. Hence for analyzes we may use moduli which, apparently, do not  appear in the measured oscillation amplitudes. In fact there are 57 groups of independent moduli, and this fact, coupled with  unitarity constraints, gives us an opportunity to find all  moduli, even nowadays we have only a limited experimental information.
 
In the present paper  we explore the possibility of extracting information on  $CP$ violation in the leptonic sector starting with the current knowledge on  moduli matrix (\ref{mns1}). We find that within the moduli intervals given by (\ref{mns1}) there are values corridors that are compatible to the existence of at least one unitary matrix, in fact one can construct a continuum of solutions, a result which could be interpreted as a signal of $CP$ violation in the leptonic sector. A clear signal of $CP$ violation will be a value $J\ne 0$ for the Jarlskog invariant, a value which can be provided by a global fit suggested in the paper. If the new fit will confirm the result, $|U_{e3}|^2<0$,  by Maltoni {\em et al.}, \cite{MSTV}, this will be equivalent with the existence of {\em new physics } in the leptonic sector. 

 The theoretical tool that will be used in the following is  unitarity of the PMNS matrix, and the constraints on physical observables generated by this property. We make full use of  phase invariance property to write the matrix under the form 
\begin{widetext}
 \begin{eqnarray}
U= \left(\begin{array}{ccc}
c_{12}c_{13}&c_{13}s_{12}&s_{13}\\
-c_{23}s_{12}e^{i \delta}-c_{12}s_{23}s_{13}&c_{12}c_{23}e^{i \delta}-s_{12}s_{23}s_{13}&s_{23}c_{13}\\
s_{12}s_{23}e^{i\delta}-c_{12}c_{23}s_{13}&-c_{12}s_{23}e^{i \delta}-s_{12}c_{23}s_{13}&c_{23}c_{13}
\end{array}\right) \label{mns}
\end{eqnarray}\end{widetext}
and the notation is  $c_{ij}=\cos \theta_{ij}$ and  $s_{ij}=\sin \theta_{ij}$,
where $\theta_{ij} $ are the mixing angles with  $ij=12, 13, 23$, and $\delta$ is the Dirac phase which encodes  $CP$ violation. 

The matrix $U$ is assumed and built as a unitary matrix, but nobody assures  that the moduli $|U_{ij}|$, the angles $\theta_{ij} $, or the phase $\delta$ extracted from experiments come from a unitary matrix. Usually it is assumed that irrespective of how the measured data are, they are compatible to the existence of a unitary matrix. A problem which was not considered in  physical literature is the property of unitary matrices to be naturally embedded  in a larger class of matrices, the set of double stochastic matrices. 

A $3\times 3$ matrix $M$ is said to be double stochastic if its entries satisfy the relations \cite{MO}
 \begin{eqnarray}\begin{array}{c}
m_{ij}\ge 0,\qquad \sum_{i=1}^3\, m_{ij}=1,\\*[2mm]
\sum_{j=1}^3\,m_{ij}=1,\qquad i,j=1,2,3\end{array}\label{ds}
\end{eqnarray}
All such matrices form a convex set, i.e. if $M_1$ and $M_2$ are double stochastic,  matrices from the set
\begin{eqnarray}
M_3=x\,M_1+(1-x)\,M_2, \,\,\, {\rm with}\quad 0\le x\le 1 \label{ds1}
 \end{eqnarray}
are double stochastic \cite{MO}. The unitary matrices are a subset of this larger set if we define the entries $m_{ij}$ of the double stochastic matrix by  relations 
\begin{eqnarray}
m_{ij}=|U_{ij}|^2\label{ds2}
\end{eqnarray}
One easily sees that relations (\ref{ds}) are satisfied because of the unitarity property
\begin{eqnarray}
U\,U^{\dagger}= U^{\dagger}\,U=I_3\label{ds3}\end{eqnarray}
where $\dagger$ denotes the hermitian conjugate matrix, and $I_3$ is the 3-dimensional unit matrix. The subset of double stochastic matrices coming from unitary matrices is known as unistochastic matrices. It is well known  in the  mathematical literature that there are double stochastic matrices that do not come from unitary matrices, see, e.g., \cite{MO} - \cite{IB}. This unpleasant feature rises  a few problems which have to be solved before trying to do a fit of the experimental data. The first one is to find the necessary and sufficient conditions the  data have to satisfy in order they should come from a unitary matrix. In other words we have to find the separation criteria between  the two sets.  After that we have to provide an algorithm for the reconstruction of $U$ from the given data. With these aims in view we have developed a formalism for expressing the necessary and sufficient conditions for a $3\times 3$ double stochastic matrix to come from a unitary one. On the other hand the  embedding relation (\ref{ds2}) suggests that a ``natural'' parametrization of $3\times 3$ unitary matrix could be that given by four independent moduli. This can be done with a proviso:  $\cos\delta$ as the function of moduli must  take physical values, i.e. $-1\le \cos\delta\le 1$. The use of  moduli is also justified by the fact that  mixing angles and   $CP$ phase are not well defined objects, they depend on the form of the unitary matrix. In this respect there are at least two popular forms: the PDG proposal \cite{CK}, and the classical proposal \cite{KM} by Kobayashi-Maskawa, and the physics has to be invariant, i.e.  not dependent  of one or another choice. The moduli are such invariants. The above statements can be formalized by the following (obvious) axiom:  irrespective of the physical processes where the moduli appear, or the phenomenological models used in analyzing the data,  the numerical moduli values should be the same.

We used the novel  formalism to develop an algorithm for the reconstruction of a unitary matrix from a double stochastic one when it is compatible to a unistochastic one. In this sense we modified the standard $\chi^2$-test to allow the implementation of all  unitarity constraints, for a better  processing of experimental error affected data.

To see that the found unitarity constraints  are necessary to be used in any phenomenological analysis we constructed an exact double stochastic matrix, from the moduli  (\ref{mns1}), which is perfectly acceptable from an experimental point of view, i.e. all the moduli are very close to the central values in  (\ref{mns1}) being within the experimental error bars, but which is not compatible with a unitary matrix. 

Last but not least we used the convexity property of double stochastic matrices to develop a method for doing statistics on unitary matrices through their squared moduli, problem which was an open one in the literature until now, see, e.g., Refs.\,\cite{BaB}-\cite{PS} .

The paper has the following structure. In Sec. II we present our phenomenological model that makes a novel use of the  unitarity property of the  PMNS matrix,  the main point being the focus on  the condition that separates  double stochastic matrices from  unitary ones. This condition is, in the same time, the consistency condition between the theoretical model, represented by the PMNS matrix, and the experimental data, as those appearing in (\ref{mns1}).
In Sec. III we present the reconstruction algorithm of unitary matrices from experimental data, when they are compatible, and show how  errors affected  data can be consistently included in the formalism. We  propose a  two terms $\chi^2$-test which has to contain a piece enforcing the fulfillment of unitarity constraints, and  another piece which  will properly   take into account  experimental data. In Sec. IV  we propose a new type of global fit, done in terms of invariant parameters, $J$ and moduli $|U_{ij}|$, which can resolve the existence of $CP$ nonconservation in the leptonic sector by using the current experimental data, or, alternatively, prove the existence of {\em new physics} in the leptonic sector. In particular we show that around central values from  data  (\ref{mns1}) there is a continuous (approximate) unitary set of matrices, all of them being consistent with  $\delta\approx \pi/2$, for a broad range of numerical $V_{e3}$ values. By changing the moduli values around one looks for unitary matrices one gets other values for $\delta$ and $V_{e3}$. The paper ends by conclusions.

\section{Phenomenological model}

By phenomenological model we will understand in the following a relationship between the entries of a double stochastic matrix, provided in general by experiments,  and the entries of a unitary matrix. The experimental data will be the numbers affected by errors entering (\ref{mns1}). We remind that in order to find allowed ranges for the moduli $|U_{ij}|$, as those given in (\ref{mns1}), 
the solar neutrino and atmospheric neutrino measurements, as well as  the reactor neutrino experiments have been used, see \cite{MSTV,FLMP,GG}.  The analyses lead to numbers  for all  moduli, but $|U_{e3}|$, from the first row and the last column. Unfortunately only  three  moduli are independent, because the unitarity gives
\begin{eqnarray}
|U_{e1}|^2+|U_{e2}|^2=|U_{\mu3}|^2+|U_{\tau3}|^2\end{eqnarray}
Taking into account the embedding (\ref{ds2}), which does not imply the unitarity fulfillment by experimental data, we will use  different notation for the theoretical quantities, and for experimental data.
 Hence in the following we assume that  the data are contained in a numerical matrix, $V,$ whose entries are positive
\begin{eqnarray}
V=\left(\begin{array}{ccc}
\vspace*{1mm}
V_{e1}&V_{e2}&V_{e3}\\
\vspace*{1mm}
V_{\mu1}&V_{\mu2}&V_{\mu3}\\
\vspace*{1mm}
V_{\tau1}&V_{\tau2}&V_{\tau3}\\
\end{array}\right)\label{exp}
\end{eqnarray}
and the current experimental data on the moduli $V_{\alpha i}, \alpha=$ $e,\,\mu,\,\tau,\, i=1,2,3,$ are summarized in  matrix (\ref{mns1}). 
From a  phenomenological point of view the main  problem to be solved  is to see if from a numerical matrix as (\ref{mns1}) one can reconstruct a unitary matrix as (\ref{mns}). For that we  define a phenomenological model as follows
\begin{eqnarray}V^2=|U|^2\label{mod}\end{eqnarray}
relation that has to be understood as working entrywise  leading to the following equations
{\small \begin{eqnarray}
V_{e1}^2&=&c^2_{12} c^2_{13},\,\, V_{e2}^2=s^2_{12}c^2_{13},\,\,V_{e3}^2=s^2_{13}\nonumber \\
 V_{\mu3}^2&=&s^2_{23} c^2_{13},\,\,
 V_{\tau3}^2=c^2_{13} c^2_{23},\nonumber\\
V_{\mu1}^2&=&s^2_{12} c^2_{23}+s^2_{13} s^2_{23} c^2_{12}+2 s_{12}s_{13}s_{23}c_{12}c_{23}\cos\delta,\nonumber\\
V_{\mu2}^2&=&c^2_{12} c^2_{23}+s^2_{12} s^2_{13} s^2_{23}-2 s_{12}s_{13}s_{23}c_{12}c_{23}\cos\delta\label{unitary},\\
V_{\tau1}^2&=&s^2_{13}c^2_{12}c^2_{23}+s^2_{12}s^2_{23}-2 s_{12}s_{13}s_{23}c_{12}c_{23}\cos\delta\nonumber,\\
V_{\tau2}^2&=&s^2_{12} s^2_{13} c^2_{23}+c^2_{12}s^2_{23} +2 s_{12}s_{13}s_{23}c_{12}c_{23}\cos\delta\nonumber
\end{eqnarray}}
together with the double stochasticity relations
{\small\begin{eqnarray}
 \sum_{i=1,2,3}| U_{\alpha i}|^2-1= \sum_{i=1,2,3} V_{\alpha i}^2-1=0, \quad \alpha=e,\mu,\nu\nonumber \\
\sum_{\alpha=e,\mu,\nu} |U_{\alpha j}|^2-1=
\sum_{\alpha=e,\mu,\nu} V_{\alpha j}^2-1=0, \quad j=1,2,3 \label{sto}
\end{eqnarray}}
The above phenomenological model is similar to that proposed by Wolfenstein, \cite{W}, i.e. it is a direct relationship between the measured values $V_{ij}$ and the theoretical parameters $s_{ij}$ and $\delta$ entering $U$. The difference between the two approaches is the explicit use of the double stochasticity properties  (\ref{sto}), and making no  approximations  on  right hand side  Eqs. (\ref{unitary}). Also important is the fact that on left hand side of Eqs. (\ref{unitary}) we have sets of nine numbers, $V_{ij}$, which  all could be obtained from experiments, and on right hand side there are {\em four} independent parameters. Hence checking the consistency of the system of Eqs. (\ref{unitary}) is a natural problem and it has to be resolved.
Since the matrix $U$ is parameterized by  four independent parameters, e.g. $s_{ij}$ and $\delta$, it is not at all obvious that the nine equations (\ref{unitary}) have a physical solution for all experimental values allowed in  (\ref{mns1}).
On the other hand we have to be aware that the parameters  $s_{ij}$ and $\delta$ are convention dependent and have no intrinsic theoretical significance, and this fact rise the problem of using only invariant functions generated by the entries of the matrix (\ref{mns}). Such functions are the Jarlskog invariant $J$ and the moduli $V_{\alpha i} $, see e.g. \cite{J1}, and this is the main reason for starting with  moduli as independent parameters in our approach.

We said before that for $n\ge 3$ there are double stochastic matrices which are not unistochastic, see, e.g. \cite{MO}-\cite{IB}, hence a novel problem arise: what are   the necessary and sufficient conditions for their separation?  In this respect there are only two  mathematical references,  \cite{YP}-\cite{N}, where theorems are given  on the  necessary and sufficient conditions for a double stochastic matrix to be also unistochastic. See also Refs. \cite{BL}-\cite{JS} for the physicists point of view.  The theorems are only   existence theorems, i.e.  they do not provide a constructive method for recovering a unitary matrix from a double stochastic one when they are compatible. However they are important since  
show  the equivalence between unitarity and the positivity of (any) area generated by the so called unitarity triangles, which is  one form of the consistency condition between double stochastic matrices and unitary ones. Here, by using the phenomenological model, Eqs.(\ref{mod})-(\ref{sto}), we provide an alternative method for the  separation criteria between the two sets, the  consequence  being a new form of the consistency conditions between the experimental data and the theoretical model, which leads eventually to a constructive algorithm for recovery of  unitary matrices from error affected data.

To see how the phenomenological model defined by the relations (\ref{mod})-(\ref{sto}) works, let  simplify the things and  assume for a moment that the relations (\ref{sto}) are exactly satisfied by the experimental data, implying the data are not affected by errors. Then it is an easy matter to find from the first five relations (\ref{unitary}) three independent ones which give  a unique solution for the $s_{ij},\, ij=12,13,23$. In other words, if the experimental numbers satisfy the relations
\begin{eqnarray}V_{e1}^2+V_{e2}^2+V_{e3}^2=1,\qquad
V_{e3}^2+V_{\mu3}^2+V_{\tau3}^2=1\end{eqnarray}
we get always a physical solution for $s_{ij}$ which is unique under the condition $0\le s_{ij}\le 1$, and depends on the three chosen independent moduli.
 Substituting this solution in the last equations one gets four equations for
 $\cos\delta$, that lead to a unique solution for it. But nobody guarantees  that the solution will satisfy the physical constraint
\begin{eqnarray}
 -1\leq\cos\delta\leq 1\label{unit1}
\end{eqnarray}
 Thus the necessary and sufficient conditions that guarantee the consistency between the data and the theoretical model are given by: 
\begin{eqnarray}
0\le s_{ij}\le 1,\quad ij=12,\,13,\,23,\quad -1\le \cos\delta \le 1\label{unt}\end{eqnarray}
The constraints (\ref{unt}) {\em together} with  relations  (\ref{unitary})-(\ref{sto}) prove the unitarity property of  data, and provide an algorithm for recovery of unitary matrices from their experimental moduli.

The embedding relations (\ref{ds2}) of unistochastic matrices into the double stochastic set  show that the number of independent parameters is the same for the two sets, i.e. in our case it is equal to four. Now we construct a double stochastic matrix by using the constraints (\ref{sto}) and the entries from the data set, (\ref{exp}). In order to simplify the  formulas form we make the notation
\begin{eqnarray} V_{e1}=a,\quad  V_{e2}=b,\quad V_{\mu1}=c,\quad  V_{\mu2}=d,\nonumber\\ V_{\mu3}=e,\quad  V_{\tau1}=g,\quad  V_{\tau2}=h,\quad  V_{\tau3}=f \label{choice}\end{eqnarray}
As we already  remarked  the  measured parameters $a,\,b,\,e, $ and $f$ are not independent  since  unitarity implies the relation
$a^2+b^2=e^2+f^2$. Thus in principle we have to chose only three of them. Our choice for the first two matrices is, $a$ and $b$, since they have the smallest  errors,  and $c$ and $e$, respectively, $d$ and $f$. The third matrix will be generated by $c,\,d,\,g,\,h$, i.e. by moduli that are not yet measured.
 If the independent parameters are $a,\,b,\,c,\,e$, the matrix $S^2=(S_{ij}^2)$, where
 \begin{eqnarray}
S=(S_{ij})=~~~~~~~~~~~~~~~~~~~~~~~~~~\qquad\qquad  \nonumber\\
\left(\begin{array}{ccc}
a&b&\sqrt{1-a^2-b^2}\\
c&\sqrt{1-c^2-e^2}&e\\
\sqrt{1-a^2-c^2}&\sqrt{c^2+e^2-b^2}&\sqrt{a^2+b^2-e^2}\end{array}\right)\label{sto1}\end{eqnarray}
is  double stochastic. Using the relations $V_{\alpha j}^2=S^2_{\alpha j}$ in Eqs.(\ref{unitary}) we find values for $s_{ij}$ and  $\cos\delta$, e.g.
\begin{widetext}
 \begin{eqnarray}
s_{12}^{(1)}&=&\frac{b}{\sqrt{a^2+b^2}},\,\, s_{13}^{(1)}=\sqrt{1-a^2-b^2},\,\,{\rm and}\,\,
s_{23}^{(1)}=\frac{e} {\sqrt{a^2+b^2}}\nonumber\\
\label{l1}\\
\cos\delta^{(1)}&=&\frac{-b^2(a^2+b^2)+c^2(a^2+b^2)^2+e^2(b^2(a^2+1)+a^2(a^2-1)) }{2\,a\,b\,e\,\sqrt{1-a^2-b^2}\,\sqrt{a^2+b^2-e^2}}\nonumber\end{eqnarray}
 In the second case when the independent parameters are $a,\,b,\,d,\,f$ one gets
\begin{eqnarray}
s_{12}^{(2)}&=&\frac{b}{\sqrt{a^2+b^2}},\,\, s_{13}^{(2)}=\sqrt{1-a^2-b^2},\,\,{\rm and}\,\,
s_{23}^{(2)}=\frac{\sqrt{a^2+b^2-f^2} }{\sqrt{a^2+b^2}}\nonumber \\
\label{l2}\\
\cos\delta^{(2)}&=&\frac{(a^2+ b^2)(b^2(1-a^2-b^2)-d^2(a^2+b^2))+f^2(a^2+b^2(a^2+b^2-1))}{2\,a\,b\,f\sqrt{1-a^2-b^2}\,\sqrt{a^2+b^2-f^2}}\nonumber
\end{eqnarray}
and, similarly, in the third case when used parameters are $c,\,d,\,g,\,h$
\begin{eqnarray}
s_{12}^{(3)}&=&\frac{\sqrt{1-d^2-h^2}}{\sqrt{2-c^2-d^2-h^2}}\, ,\quad s_{13}^{(3)}=\sqrt{c^2+d^2+g^2+h^2-1},\quad  s_{23}^{(3)}=\frac{\sqrt{1-c^2-d^2}}{\sqrt{2-c^2-d^2-h^2}\label{l3}}\end{eqnarray}
 \begin{eqnarray}
\cos\delta^{(3)}=\left(c^2(1-c^2)-d^2(1-d^2)-g^2(1-g^2)+h^2(1-h^2)+(d^2 g^2-c^2 h^2)(c^2+d^2+g^2+h^2-2)\right)/ \label{l4} \\ \left( 2\,\sqrt{1-c^2-d^2}\,\sqrt{1-c^2-g^2}\,\sqrt{1-d^2-h^2}\,\sqrt{1-g^2-h^2}\,\sqrt{c^2+d^2+g^2+h^2-1}\right)\nonumber
\end{eqnarray}
In the first two cases the mixing angles $s_{ij}$ depend only on three parameters, such that the  only constraint on  all four parameters is given by 
$-1\le \cos\delta\le1$ which we  write it in the form
\[1-\cos^2\delta=\sin^2\delta \ge 0\]
By using  the form  (\ref{l1}) for $\cos\delta$ the above condition is equivalent to
{\small
\begin{eqnarray}
&&\sin^2\delta=~~~~~~~~~~~~~~~~~~~~~~~~~~~~~\qquad\qquad\qquad \qquad\qquad    ~~~~~~~~~~~~~~~~~~~~~~~~~~~~~~~~~~~~~~~~~~~~~~~~~~~~~~~~~~~~~~\nonumber\\
&&\frac{-(a^2+b^2)^2\left(b^2(b^2-2\,c^2(a^2+b^2))+c^4(a^2+b^2)^2+e^4(1-a^2)^2+2\,e^2\,b^2(a^2-1)+ 2\,c^2\,e^2(a^2(a^2-1)+b^2(1+a^2)\right)}{
4\,a^2\,b^2\,e^2(1-a^2-b^2)(a^2+b^2-e^2)} \ge 0\qquad \label{tri}\end{eqnarray}
On the other hand Branco and Lavoura, \cite{BL}, and, respectively, Jarlskog and Stora, \cite{JS}, found a geometric invariant, the area of any unitarity triangle,  as a measure of  $CP$ violation. This area has to be a positive number if the $CP$ symmetry is violated. If $l_i$,\,\,i=1,2,3, denote the lengths of the triangle sides, from (\ref{sto1}) one gets
\begin{eqnarray} l_1=a\,b,\quad l_2=c\,\sqrt{1-c^2-e^2},\quad 
l_3=\sqrt{1-a^2-c^2}\sqrt{c^2+e^2-b^2}\end{eqnarray}
One  makes use of Heron's formula 
$A=\sqrt{p(p-l_1)(p-l_2)(p-l_3)}$,
where $p=(l_1+l_2+l_3)/2$ is the semiperimeter, for getting  area $A$, and the reality   condition $A^2\ge 0$ has the form
{\small
\begin{eqnarray}16\,A^2&=& -(b^2(b^2-2\,c^2(a^2+b^2))+c^4(a^2+b^2)^2+e^4(1-a^2)^2+2\,e^2\,b^2(a^2-1)+ 2\,c^2\,e^2(a^2(a^2-1)+b^2(1+a^2)))\ge 0\label{tri1}
\end{eqnarray}}}\end{widetext}
Looking at equations (\ref{tri}) and (\ref{tri1}) we see that they are equivalent, the supplementary factors appearing in (\ref{tri}) being positive for a double stochastic matrix. Hence the condition  (\ref{unt}) on $\cos\delta$ is a novel separation criterion between the unistochastic matrices and the double stochastic ones.
The separation border is given equivalently by $A=0$, or $\cos^2\delta= 1$.

Equivalent conditions to (\ref{tri}) were given in \cite{BL}-\cite{H}, where  positivity of $\sin^2\delta\ge 0$  was mainly used to constrain the four implied moduli. One aim of the paper is to put together all that information for devising a fitting model which will use all the possible constraints coming from unitarity. The idea behind it is to use the non uniqueness of the set of four independent moduli used to parameterize a unitary matrix which leads to many relationships between them and which over constrain the  error  affected experimental data.  A consequence will be that unitarity implies a very precise tuning between all  moduli, and we could get information on the physical relevant parameters even from a limited  experimental information.

From (\ref{l1})-(\ref{l3}) we see that the expressions defining $s_{ij}$ and  $\delta$ are quite different. The interesting thing is that all the possible forms for $s_{ij}$ and  $\cos\delta$ take the same numerical values when the numerical matrix is double stochastic. Let us consider the following matrix
\begin{eqnarray}
S&=&\left(\begin{array}{ccc}
\frac{167}{200}&\frac{11}{20}&\frac{\sqrt{11}}{200}\\*[1mm]
\frac{8}{25}&\frac{\sqrt{237}}{25}&\frac{18}{25}\\*[1mm]
\frac{1}{40}\sqrt{\frac{1603}{5}}&\frac{\sqrt{3183}}{100}&\frac{\sqrt{19253}}{200}\\*[1mm]\end{array}\right)\label{num}\\
 &\approx& \left(\begin{array}{lll}
0.835&0.55&0.0166\\
0.32&0.6158&0.72\\
0.4476&0.5642&0.6938\end{array}\right)\nonumber
\end{eqnarray}
whose entries are within the error bars from (\ref{mns1}), and  the matrix $S^2= (S_{ij}^2)$ is a double stochastic matrix. The numbers entering it have been  chosen such that the measured parameters, $V_{e1}$,\,  $V_{e2}$,\,  $V_{\mu3}$ and   $V_{\tau3},$ should be around the central values from (\ref{mns1}). For this matrix we get
\begin{eqnarray} s_{12}^{(i)}&=&\frac{110}{\sqrt{39989}}\approx 0.55,\quad s_{13}^{(i)}=\frac{\sqrt{11}}{200}\approx 0.0166,\nonumber\\
 s_{23}^{(i)}&=&\frac{144}{\sqrt{39989}}\approx 0.72,\,\quad i=1,2,3\label{num1}\end{eqnarray}
\begin{eqnarray}
\cos\delta^{(i)}=-\frac{5419565144}{2066625\sqrt{211783}}\approx -5.698, \,\, i=1,2,3\label{num2}\end{eqnarray}
The numerical results show that all the three angles $s_{ij}$ are physical, but $\delta$ is not, i.e.
 the data matrix $S$, Eq.(\ref{num}), does not come from a unitary matrix.
Computing the triangle area we find
\begin{eqnarray}A=0.0107\,i\end{eqnarray}
i.e. an imaginary number which sends the same signal, namely,   the above data (\ref{num}) are not compatible to the existence of a unitary matrix.
 
 Thus the new unitarity constraint $-1\le\cos\delta\le 1$ puts non trivial constraints on the experimental data. From an experimental point of view the above matrix is perfectly acceptable, from a theoretical point of view it is not; and the non-unitarity of the data  can be checked  by the above method.

In  the case when the numerical matrix (\ref{mns1}) is  double stochastic, as the above matrix $S$, $s_{ij}$ and $\cos\delta$  do not depend on the chosen four independent moduli used to define it. If  the   modulus of  $\cos\delta$ is outside the interval $(-1,1)$ the data are not compatible to the theoretical model and the story ends here. If  $\cos\delta$ takes values within the interval $(-1,1)$,  the data are compatible with the theoretical model, and the values for $\delta$  and $s_{ij}$ necessary for the recovery of the unitary matrix are easily obtained, e.g. from the relations (\ref{l1}), and by introducing them in Eq.(\ref{mns}) we explicitly recover the unitary matrix. 

 From almost any numerical experimental data as (\ref{mns1}) we can form a doubly stochastic one by choosing at our will four independent parameters, the other five being completely determined by equations similar to (\ref{sto1}). In this way we obtained the equations  (\ref{l1})-(\ref{l3}) for $\cos\delta$. By using  numerical information from  matrix (\ref{mns1}) for $a,\,b,\,c,\,d,\,e,\,g,\,h$ and $f$, we get
\begin{eqnarray}
\cos\delta^{(1)}_+&=&-1.1\,i,\,\,\cos\nonumber\delta^{(1)}_c=-0.553,\nonumber\\*[2mm]cos\delta^{(1)}_-&=&-0.985,\,\,
\cos\delta^{(2)}_+=0.94 i\,\, \nonumber\\*[2mm]
\cos\delta^{(2)}_c&=&0.118,\,\,\cos\delta^{(2)}_-=0.737\label{l10}\\*[2mm]
\cos\delta^{(3)}_+&=&0.05 i,~\,\,\cos\delta^{(3)}_c=0.018 i,\nonumber\\*[2mm]
\cos\delta^{(3)}_{-}& =& 0.0027 i,\nonumber
\end{eqnarray}
where  $+,\,c,\,-$  denote  values of  $\cos\delta$  function obtained from  central values$+\sigma$, central values, and, respectively,  central values$-\sigma$. The corresponding values for $\cos\delta$ on the three rows are different because the numerical values for  parameters $a,\,b,\,c,\,  e$, respectively, $d,f,g,h$ do not come from the same doubly stochastic matrix,  e.g. $e\ne\sqrt{1-c^2-d^2}$, or numerically $0.7\ne 0.737$. The above numerical computations send a clear signal:  unitarity of the PMNS matrix is a property shared by all its entries implying strong correlations between the numerical values taken by $V_{ij}$, and this unitarity feature has to be well understood by people doing phenomenology. The border of the physical region, $A=0$, or $\cos\delta=\pm 1$, can be used to find maximal intervals for the unmeasured moduli. If we consider $V_{e1},\,V_{e2},\, {\rm and}\,\, V_{\mu3}\,\,{\rm or}\,\, V_{\tau3}$ fixed to  the central values 
 one  gets  $V_{\mu1} \in (0.31,\,0.45)$ which represent 40\% from the 3 $\sigma$ interval given by (\ref{mns1}), and respectively, $V_{\mu2} \in (0.53,\,0.64)$, i.e. about 35\% from the 3 $\sigma$ interval. If one computes now the same quantities by using the non unitary double stochastic matrix (\ref{num}) one finds $V_{\mu1} \in (0.37,\,0.39)$, and respectively, $V_{\mu2} \in (0.57,\,0.59)$, i.e. intervals up to 7 times shorter than in the previous case. In conclusion the constraints are stronger for matrices which are  far from unitary ones. The previous computations show also how one can modify the numerical matrix (\ref{num}) for obtaining a unitary compatible one.

Taking into account that more often than not  unitarity constraints are not satisfied by  experimental data as the numerical results (\ref{l10}) show we have to devise a fitting model by supplementing the usual $\chi^2$-test with a separate piece that should implement the fulfillment of  unitarity constraints,  if we want to get reliable  values for the interesting  physical  parameters.

\section{Reconstruction of unitary matrices from data  }

We have seen in the previous section that by using  numerical values from  neutrino data, Eq.(\ref{mns1}), one finds values for 
$\cos\delta$ outside
the physical range $[-1,1]$,  and generically all  numerical values are different. The expressions for $\cos\delta$ are provided by 
the last four relations (\ref{unitary}), and these formulas have to give the same number when comparing theory with experiment, by supposing the data come from a unitary matrix. Their explicit form depends on the four independent moduli   used  to parameterize the data, see e.g. (\ref{l1})-(\ref{l3}), and since there are 57 such independent groups one gets  165 different expressions for  $\cos\delta$.

Hence if  data are compatible to the existence of a unitary matrix the  mixing angles $s_{ij}^{(m)}$ coming from all 57 groups, and all  165 phases $\delta^{(m)},$ expressed as functions of $V_{\alpha i} $, have to be (approximately) equal, and these are  the most general necessary  conditions for unitarity; they  can be written as
\begin{eqnarray}&& 0\le s_{ij}^{(m)}\le 1, \,\, s_{ij}^{(m)}=s_{ij}^{(n)},\,\,m, n=1,\dots,57,\label{uni}\\ &&\cos\delta^{(i)}=\cos\delta^{(j)},\,i,j=1,\dots,165\nonumber\end{eqnarray}
The  above relations are also satisfied  by  double stochastic matrices, as the numerical relations from Eqs.(\ref{num1}) and (\ref{num2}) show, and the condition that separates  unitary matrices from   double stochastic ones is given by  relation (\ref{unit1}), i.e. $-1\le \cos\delta^{(i)}\le 1$.

As a warning, the content of relations (\ref{uni}) can be summarized as follows:   unitarity implies strong correlations between  all numerical values $V_{\alpha j}$.

Now we present the necessary input that have to be taken into account when doing a fit, the experimental data are considered  moduli of a unitary matrix. It has to include as a separate piece the conditions for a  complete implementation of  unitarity. For that  we define a test function that has to take into account the double
stochasticity property expressed by the 
conditions (\ref{sto}) and the fact that in general the numerical values of data are such that
  $\cos\delta$ depends on the choice of the four independent moduli and could take values outside the physical range. The proposal for the first piece is

\begin{eqnarray}
\chi^2_1&=&\sum_{i < j}(\cos\delta^{(i)} -\cos\delta^{(j)})^2+\sum_{\alpha=e,\mu,\tau}\left(
\sum_{i=1,2,3}V_{\alpha i}^2-1\right)^2\nonumber\\
&+&\sum_{\alpha=e,\mu,\tau}\left(
\sum_{i=1,2,3}V_{\alpha j}^2-1\right)^2,\,\,\,\,-1\le\cos\delta^{(i)}\le 1\label{chi1} 
\end{eqnarray}

This proposal was made in \cite{Di} for the case of Cabibbo-Kobayashy-Maskawa (CKM) quark  matrix, and is our  contribution to the existing methods of reconstructing   unitary matrices from experimental data, when  data are compatible with their existence;  it expresses the full content of unitarity. We stress again that  both  conditions have to be fulfilled: $\chi^2_1$ has to take small values and $\cos\delta$ values have to be physically acceptable, since   the  relation $\chi^2_1\equiv 0$ holds true for both double stochastic and unitary matrices. When the experimental data are those from the matrix (\ref{mns1}) the second piece 
 has the form
\begin{eqnarray}
\chi^2_2=\sum_{i=1,2,3}\,\,\sum_{\alpha=e,\mu,\tau}\left(\frac{V_{\alpha i}-\widetilde{V}_{\alpha i}}{\sigma_{\alpha i}}\right)^2\label{chi2} 
\end{eqnarray}
where  $V_{\alpha i}=|U_{\alpha i}|$ are the minimizing parameters, and $\widetilde{V}_{\alpha i}$ is  the numerical matrix that describes the experimental data as in (\ref{mns1}), and $\sigma$ is the errors matrix  associated to $\widetilde{V}_{\alpha i}$; in our case $\widetilde{ V}_{e1} =0.835$, $\sigma_{e1}=0.045$, and so on. 

  Thus the simplest test function has the form
\begin{eqnarray} \chi^2=\chi^2_1+\chi^2_2\label{chi3} 
\end{eqnarray}
Of course  unitarity constraints (\ref{chi1}) could be implemented by penalty functions, and the  concrete form of (\ref{chi3}) could depend on the quality of the fit, which finally will  choose between the form (\ref{chi1}), or that with penalty functions. 
The formula (\ref{chi3}) provides  a least squares method which  is similar to that used by the {\em BABAR} Collaboration \cite{BaB}, the difference being the term $\chi^2_1$ that implements the theoretical model (\ref{mns}).

The above form can be easily modified if we have other experimental information for some functions depending on $s_{ij}$ and $\delta$, or on $V_{\alpha i}$. The primary data in neutrino physics are the transition probabilities between two neutrino flavors $\nu_{\alpha}\rightarrow \nu_{\beta}$, $\alpha,\,\beta= e,\,\mu,\,\tau$ at a distance $L$, whose generic form is  \cite{AS}
\begin{eqnarray}
P(\nu_{\alpha} \rightarrow \nu_{\beta}; E,L)=\sum_{i\,j}U_{\beta i}U_{\beta j}^* U_{\alpha i}^* U_{\alpha j}\,{\rm exp}(-i \,\Delta_{ij}^2)\nonumber\\
=\sum_{i,\, j} \Re(W_{\alpha\beta}^{ji})\,\cos\Delta_{ij}^2 +\sum_{i,\,j} \Im(W_{\alpha\beta}^{ji})\,\sin\Delta_{ij}^2\label{osc}~~~\end{eqnarray}
where  
 \begin{eqnarray}W_{\alpha\beta}^{ji}=U_{\alpha j}U_{\beta j}^* U_{\alpha i}^* U_{\beta i},\,\, \Delta_{ij}^2=(m_i^2-m_j^2)L/2 E_{\nu}\,\, \end{eqnarray}  and  $\alpha=e,\,\mu,\,\nu,\,\,i,j=1,2,3$.
From \cite{JS}, see also \cite{Cecil}, we know that the  imaginary part has the form
\begin{eqnarray}
\Im (U_{\alpha j} U_{\beta j}^*
U_{\alpha i}^* U_{\beta i})= J\,\sum_{\gamma,\, l} \epsilon_{\alpha \beta \gamma}\,\epsilon_{ j i l}\end{eqnarray}
with $J$ the Jarlskog invariant. On sees that the parameter involving the $CP$ violation appears  in  transition probabilities, hence it can be directly determined from the
experimental data. By using the above relation one gets
\begin{eqnarray}
\sum_{i,\,j} \Im(U_{\alpha j} U_{\beta j}^*
U_{\alpha i}^* U_{\beta i})\,\sin\Delta_{ji}^2
=\nonumber\\\pm 2\,J(\sin\Delta_{12}^2+\sin\Delta_{31}^2+\sin\Delta_{23}^2)\end{eqnarray}
The term coming from the real part $\Re(W_{\alpha \beta}^{ji} ) $ depends explicitly on the flavors $\alpha \,\,{\rm and}\,\,\beta $. When $\alpha=\mu$ and
 $\beta=e$ one gets
\begin{eqnarray}
&&\sum_{i,\, j}\Re(W_{\mu e}^{ji} )\,\cos\Delta_{jk}^2=~~~~~~~~~~~~\\ 
&&R_{12}\,\cos\Delta_{12}^2 +R_{13}\,\cos\Delta_{31}^2+R_{23}\,\cos\Delta_{23}^2
\nonumber\end{eqnarray}
where
\begin{eqnarray}\begin{array}{lll}
R_{12}=&U_{e1}\,U_{e2}\,\Re(U_{\mu 1}\,U_{\mu 2}^*)\\*[2mm]
~~~~~=& -(c_{12}^2 c_{13}^2 s_{12}^2(c_{23}^2-s_{13}^2 s_{23}^2)\\*[2mm]
&+ c_{12} c_{13}^2 c_{23} s_{12} s_{13}s_{23}(c_{12}^2-s_{12}^2\cos\delta)\\*[2mm]
R_{13}=&U_{e1}\,U_{e3}\,U_{\mu 3} \,\Re(U_{\mu 1})\\*[2mm]
~~~~~=& -c_{12}c_{13}^2s_{13}s_{23}(c_{12}s_{13}s_{23}+c_{23}s_{12}\cos\delta \\*[2mm]
R_{23}=&U_{e2}\,U_{e3}\,U_{\mu 3}\Re(U_{\mu 2})\\*[2mm]
~~~~~= &c_{13}^2s_{12}s_{13}s_{23}(c_{12}c_{23}\cos\delta-s_{12}s_{13}s_{23})\end{array}\label{real1}\end{eqnarray}
The above formulas have been obtained by using the standard form (\ref{mns}) of the PMNS matrix. We use now the phenomenological model (\ref{real1}) 
to write $R_{ij}$ in terms of invariant  quantities, i.e. moduli.
 One gets
\begin{eqnarray}
R_{12}&=&\frac{1}{2}(-b^2-a^2 c^2+b^2 c^2+e^2-a^2 e^2)\nonumber\\
R_{13}&=&\frac{1}{2}(b^2-a^2 c^2-b^2c^2-e^2+a^2 e^2)\label{real2}\\
R_{23}&=&\frac{1}{2}(-b^2+a^2 c^2+b^2 c^2-e^2+a^2 e^2+2 b^2 e^2)\nonumber\end{eqnarray}

Similar formulas have to be obtained for all  amplitudes entering the experimental data. For example  the three-flavor survival probability $P_{ee}$ in vacuum \cite{MSTV} is given by
{\small\begin{eqnarray}
P_{ee}&=&1-\frac{1}{2} \sin^2 2\theta_{13}-\cos^4\theta_{13}\,\sin^2 2\theta_{12}\,\sin^2\Delta_{21}^2\label{real3}\\
&=&1-2(a^2+b^2)+2(a^2+b^2)^2-4\,a^2 b^2\sin^2\Delta_{21}^2\nonumber
\end{eqnarray}}

Hence the transition probabilites, (\ref{real1})-(\ref{real3}), and those similar to them,  depends on eight parameters: $a,\,b,\, c,\,e,\, J,$ $ \Delta_{12},\,\Delta_{13}, \Delta_{23}$ which can be used to fit the data. In the usual assumption that $U$ is unitary only six  are independent, for example $a,\,b,\, c,\, J,$ $ \Delta_{12},\,\Delta_{13}$. That means that we may eliminate  modulus $e$. For that we  use the relation  $J= 2\, A$ where, $J$ is the Jarlskog invariant, and $A$ is the area of (any) unitarity triangle. One gets

\begin{widetext}\begin{eqnarray}
e^2=\frac{b^2(1-a^2-c^2)+a^2c^2(1-a^2-b^2)}{(1-a^2)^2}+
 \frac{2 \epsilon \sqrt{ a^2 b^2 c^2(1-a^2-c^2)(1-a^2-b^2)-(1-a^2)^2 J^2}}{(1-a^2)^2}~~\label{real4}\end{eqnarray}\end{widetext}

\noindent
 where $\epsilon=\pm 1$. If the matrix is unitary then the following condition has to be satisfied $0\le e^2\le1 $. The second parameter is eliminated from the relation
\[\Delta_{12}^2+\Delta_{23}^2+\Delta_{31}^2=0\]
 If we use the the phenomenological models (\ref{l2}) and (\ref{l3}), i.e. if we take as parameters   four independent moduli, we find form different terms for all $R_{ij}$ appearing in formulae (\ref{real1}), see Eqs. (\ref{real2}) and (\ref{real3}). Thus $\chi^2_2$ could have the form
\begin{eqnarray}
\chi^2_2=\sum_{\alpha}\,\,\sum_{i}\,\,\sum_{j}\left(\frac{P_i(\nu_{\alpha\rightarrow \nu_{\beta}})-\widetilde{P^j(\nu_{\alpha}\rightarrow \nu_{\beta})}}{\sigma_{P^j}}\right)^2\label{newchi2} 
\end{eqnarray}
where $\alpha$ runs over all  transitions probabilities involved in the fit, $i$ runs over the sets of four independent moduli, such as those given by  relations (\ref{l1})-(\ref{l3}), which give equivalent but  form different expressions for the theoretical probabilities $P(\nu_{\alpha\rightarrow \nu_{\beta}}) $, and $j$ runs over the number of experimental points for each transition probability. By doing a fit with the form (\ref{chi3}) where $\chi^2_2$ is given by (\ref{newchi2}) one gets values for  moduli $V_{\alpha i}$, $\Delta_{ij}$ and $J$. 

The fit could provide a few (approximate) unitary matrices compatible with the data. In such a case the double stochastic matrices obtained from  fit could be used to do statistics on  unitary matrices.
At this point the embedding property of unitary matrices into the convex set of double stochastic matrices  is essential.
Indeed if $U_1,\,U_2,\,\dots\,U_n$ are unitary matrices, then
\begin{eqnarray}
M^2&=&\sum_{i=1}^{i=n} \,x_i \cdot |U_i|^2, \qquad \sum_{i=1}^{i=n}x_i=1,~~~~ \label{db} \\
&& \,\, 0\le x_i \le 1, \qquad i=1,\dots,n\nonumber\end{eqnarray}
is double stochastic. The above relation shows that the statistics on unitary matrices has to be done through the statistics on double stochastic matrices, i.e. the relevant quantities for doing statistics on unitary matrices are the squared moduli, $|U_{\alpha j}|^2.$  This property allows us to define correctly   the mean value, $\langle M \rangle $, and the error matrix,  $\sigma_M$, for a set of doubly stochastic matrices, coming  from a set of approximate unitary matrices,  as follows
\begin{eqnarray}
\langle M \rangle&=&\sqrt{\left(\sum_{k=1}^{k=n} \, |U_k|^2\right)/ n}\\*[3mm] 
\sigma_M&=&\sqrt{\left(\sum_{k=1}^{k=n} \, |U_k|^4\right)/ n\,\, -\,\langle M\rangle ^4}
\label{db1}\end{eqnarray}
 If the moduli of the mean value matrix $\langle M \rangle$ obtained in this way are compatible with a unitary matrix, or  not too far  from moduli coming from a unitary one, one can reconstruct from it an (approximate) unitary matrix, and in the following we use this technique to reconstruct unitary matrices compatible to the data, and  for obtaining information on relevant physical parameters.

\section{Global fits}

By global fit we  understand a fit that takes into account all  available experimental data, and (all)  unitarity constraints expressed by invariant functions; on this last requirement see, e.g.,  Ref.\,\cite{CJ1}, the aim in view being a good estimation of mixing and $CP$ violation in neutrino physics. 
For doing that we use the simple ansatz (\ref{chi1})-(\ref{chi3}) in conjunction with  data (\ref{mns1}) to obtain  qualitative results about $J$, $\delta$ and/or  $\widetilde{V}_{e3}$. The last parameter was considered free within the interval  $\widetilde{V}_{e3}\in (0.02-0.4)$ and its error  was taken  $\sigma_{e3}=\widetilde{V}_{e3}$.
Around the central values from (\ref{mns1}) we obtained many approximate double stochastic matrices, i.e. values for all   moduli $V_{\alpha i}$, and by doing statistics with formulae (\ref{db1}) the following 
 mean value matrix and  associated error matrix  were gotten:
\begin{eqnarray}
M_1  = \left(\begin{array}{ccc}
  0.835392& 0.537838& 0.114933\\
  0.381844& 0.585045& 0.715381\\
       0.395402& 0.607056& 0.689253  \end{array}\right)\label{m1}
\end{eqnarray}

\begin{eqnarray}
\sigma_{M_1}= \left(\begin{array}{ccc}
  0.004241& 0.00467& 0.009417\\
  0.001008& 0.00089& 
        0.000263\\
  0.003287& 0.005701& 0.009052\end{array}\right)\label{s1}
\end{eqnarray}
and the unitary matrix corresponding to $M_1$ is
\begin{widetext}
\begin{eqnarray}
U_1=\left(\begin{array}{ccc}
0.835176&0.537838&0.114933\\
-0.0697446-0.375651\,i&-0.0445706+0.583326\,i&0.715381\\
-0.0668811+0.38991\,i&-0.0434268-0.605467\, i&0.689253
\end{array}\right)\label{u1}
\end{eqnarray}\end{widetext}
If we define two matrices, let us say, by the relation
\begin{eqnarray}M_{\pm}=\sqrt{M^2_1\pm 3\,\sigma^2_{M_1}}\label{db2}\end{eqnarray}
we get the following prediction for $CP$ violating phase and unitarity triangle area
\[\delta=(89.98^{+0.22}_{-0.26})^{\circ},\quad A=(127.44\pm 0.01)\times 10^{-4}\]
corresponding to a $3\, \sigma$ level, and so on.

The results (\ref{m1}) and (\ref{s1}) strongly depend on the moduli values around one looks for a minimum. Changing these values within the error bars in (\ref{mns1}) one gets other matrices and other values for $\delta$. For example the matrix 
\begin{eqnarray}
M_2  = \left(\begin{array}{ccc}
  0.851242& 0.523442& 0.0373568\\
  0.34007& 0.578624& 0.741314\\
       0.399674& 0.625462& 0.670118  \end{array}\right)\label{m2}
\end{eqnarray}
leads to a value $\delta=(120.01\pm 0.16)^{\circ}$, $J\approx 2.\times 10^{-4}$, and so on. This happens since the errors entering (\ref{mns1}) are too large, such that for many choices of fixed three moduli,   $\delta$ is practically unconstrained, i.e. there is a continuum of unitary matrices  with $\cos\delta$ varying in many cases between $(-1,1)$.

The  result (\ref{m1}) is ``confirmed'' by  a ``standard'' fit when one uses  formulas (\ref{unitary}), without any approximations coming from small parameters, etc., and one takes as independent parameters $\theta_{ij}$ and $ \delta$. The term containing $\widetilde{V}_{e3}$ was not introduced in the fit. One gets easily
\begin{eqnarray} \theta_{12}&=&0.572332,\,\,\theta _{13}=0.088198,\,\,\theta _{23}=0.796547, \nonumber\\
 \delta&=&90.51^{\circ}\label{fit1},\,\, A=9.9\times 10^{-3}\end{eqnarray}
where the values for $\theta_{ij}$ are given in radians, and $\chi^2=8.7\times 10^{-2}$, where from one gets   the following double stochastic matrix 
\begin{eqnarray}
M_3  = \left(\begin{array}{ccc}
  0.837373& 0.539489& 0.0880842\\
  0.381844& 0.589052& 0.712167\\
       0.391116& 0.601639& 0.696462  \end{array}\right)\label{m3}
\end{eqnarray}
The corresponding unitary matrix is given by
\begin{widetext}\begin{eqnarray}
U_2=\left(\begin{array}{ccc}
0.837373&0.539489&0.0880842\\
-0.0495463-0.378656\,i&-0.0393741+0.587735\,i&0.712167\\
-0.0552421+0.387195\,i&-0.0279691-0.600988\, i&0.696462
\end{array}\right)\label{u3}
\end{eqnarray}\end{widetext}
The remarkable fact is that the fit provides a numerical value for $V_{e3}$ which could not be too far from the true one, although the global fit \cite{MSTV} failed to find it. To estimate the errors to the above parameters we used  the moduli matrices $\sqrt{V^2- 2\,\sigma^2}$ and $\sqrt{V^2+4\,\sigma^2}$, where $V$ and $\sigma$ are those given by (\ref{mns1}), and doing again the fit we got 
$\chi^2_{2\sigma}=1.32$, and, respectively,  $\chi^2_{4\sigma}=1.22$, where  the following prediction was obtained
\begin{eqnarray}\delta=(90.51^{+2.29}_{-1.24})^{\circ}, \quad A=(9.2^{+0.7}_{-2.7})\times 10^{-3}\end{eqnarray}

To see the difference between statistics done by using (\ref{db1}) and (\ref{db2}) and the naive one, we used the central matrix (\ref{mns1}), and the corresponding error matrix $\sigma$ obtaining    $V_c +\sigma$,
to find
\begin{eqnarray}\theta _{12}&=&0.57996,\quad\theta_{13}=0.00054,\nonumber\\\theta_{23}&=&0.797221,\quad \delta=79.54^{\circ}\label{fit2} \end{eqnarray}
results which are equivalent with  the  moduli matrix 
\begin{eqnarray}
W  = \left(\begin{array}{ccc}
  0.836484& 0.547991& 0.00054\\
  0.382939& 0.584411& 0.715417\\
       0.391984& 0.598473& 0.698697  \end{array}\right)\label{w}
\end{eqnarray}
This matrix is not compatible with a unitary matrix if we use our criteria. For example one gets 
$\cos\delta^{(3)}=-0.1046\,i$,
and  $A =  9.0 \times 10^{-5}\,i$. In fact there are 20 $\cos\delta$ from a total of 165 which take imaginary values, so, in principle, the problem cannot be neglected.  On the other hand   one finds $\chi^2=6.4 $ for 4 d.o.f., which is not too bad. Taking into account that the double stochasticity property (unitarity) is only weakly violated, we find from (\ref{fit2}) a conservative estimate
$\delta\ge 70^{\circ}$.  
 Doing a similar fit for $V-\sigma $ one gets $\chi^2=8.1 $ for 4 d.o.f., and $\delta=91.81^{\circ}$.

 Of course in true fits one does not work with quantities such as $V\pm \sigma $, and they were given as pedagogical examples. They show  that  it is possible when using as independent parameters $\theta_{ij}$ and $\delta$ to  obtain physical values for all of them, although the area of unitarity triangle could take imaginary values.   The above computations show  that  $s_{ij}$ and $\delta$ are convention dependent, even if their numerical values are physical, and in order to minimize this  bias we expressed these parameters in terms of the moduli $V_{\alpha i}$ which have to be the same irrespective of the used convention for the unitary matrix (\ref{mns}), and  we required the fulfillment of all  constraints (\ref{uni}). The second lesson is that one could get values for $\delta$ even if  moduli where $\delta$ enters are not yet measured, and this happens since unitarity over constrains the data implying many relationships between the moduli. Of course a negative attitude to such results is compliant so long as the subtleties implied by unitarity are not fully understood.

In the following we want to stress the non uniqueness  of unitary matrices compatible with the error bars in (\ref{mns1}). For example,   one can find  also a unitary matrix which lead to $\delta =(75.54\pm 0.11)^{\circ} $; its moduli are 
\begin{eqnarray}
M_4  =
 \left(\begin{array}{ccc}
  0.813246& 0.575234& 0.087959\\
0.453777& 0.614523& 
      0.645328\\
0.364304& 0.539877& 0.758825
  \end{array}\right)\label{m4}
\end{eqnarray}

With the matrices $M_i,\,\, i=2,3,4$, by using the property of double stochastic matrices we find a continuum of (approximate) unitary matrices whose squared moduli are given by 
\begin{eqnarray}
M^2=x\,M_2^2+y\,M_3+(1-x-y)\,M_4^2, \,\, 0\le x, y \le 1~~ 
\end{eqnarray}
Doing statistics on $M_i,\,i=2,3,4$, with formulae (\ref{db1}) one gets
\begin{eqnarray}
M_5  =  \left(\begin{array}{ccc}
  0.834101& 0.546484& 0.0750359\\
0.394701& 0.594258& 
      0.700756\\
0.385327& 0.590096& 0.694424
  \end{array}\right)\label{m5}
\end{eqnarray}

\begin{eqnarray}
\sigma_{M_5}= \left(\begin{array}{ccc}
  0.026107&0.023843 &0.002995\\
  0.037521&0.0180207& 
       0.05552  \\
  0.01148& 0.041863& 0.06711\end{array}\right)\label{s2}
\end{eqnarray}
With  matrix $M$ one can obtain any value for $\delta\in (75^{\circ}-120^{\circ})$, and, respectively, $A\in(2\times 10^{-4}-1.6\times10^{-3})$. The prediction
 at the symmetric point $x=y=1/3$ obtained from $M_5$  is
\[\delta=(88.13\,^{+ 14.05}_{- 10.45})^{\circ}, \quad A=(11.1\pm 0.3)\times 10^{-4}\]
at one sigma level given by  $\sigma_{M_5} $.

In the following we want to test the dependence of  moduli matrix on the  form of the unitary matrix $U$. For that we use the Kobayashi-Maskawa form \cite{KM}
{\small\begin{eqnarray}
U=
\left(\begin{array}{ccc}
c_{1}&c_{3} s_{1}&s_{1} s_{3} \\
c_{2} s_{1}&-c_{1} c_{2} c_{3} +  s_{2}s_{3}\, e^{i \varphi}&-c_{1}c_{2} s_{3}-c_{3}s_{2}\, e^{i\varphi }\\
s_{1}s_{2}&-c_{1} s_{2}c_{3}- c_{2} s_{3}\, e^{i\varphi}&-c_{1} s_{2} s_{3}+c_{2}c_{3}\, e^{i\varphi }
\end{array}\right) \label{mns2}
\end{eqnarray}}
where $c_i=\cos\theta_i$ and $s_i=\sin\theta_i$, and $\varphi$ is the Dirac phase. Of course selecting one or another form has no theoretical significance because all  choices are mathematically equivalent; however a clever choice may shed some light on important issues. Such  issues could be the  $|U_{e3}|$ magnitude and the $CP$ phase $\delta,\,(\varphi)$, and  the above form (\ref{mns2}) could be more helpful.

Doing a fit by using the relations (\ref{unitary}) and the new parameters $\theta_i$ one gets the moduli matrix (\ref{m3}) with the same digits. In fact we get
\begin{eqnarray}
\theta_1&=&0.578337,\quad \theta_2=0.797342,\\ \nonumber\theta_3&=&0.161845,\quad \varphi=90.15^{\circ}\end{eqnarray}
that confirm the previous results, the  only change being the value for $\theta_3$ which is brought, in principle, to a measurable value.

A few conclusions of the above numerical fits could be: (a)  there are large corridors for  moduli around the central values in  (\ref{mns1}) where  data are compatible to the existence of a unitary matrix with $\delta\ne 0,\,\pi$, and by consequence compatible with a $C$P violation; (b) the real parts of the complex entries in unitary matrices are  smaller by an order of magnitude than the imaginary parts, see e.g. matrix $U_1$, but their neglect completely spoils the unitarity property; (c) by consequence,  reliable results can be obtained then and only then when   all  theoretical formulas used in  fits are exact, {\em not approximate}.

We are aware that the ``experimental'' data (\ref{mns1}) are not  numbers obtained directly from experiments. In the following we  suggest  a new kind of global fit able to resolve the existence or the lack of the $CP$ nonconservation in the leptonic sector by measuring directly the Jarlskog invariant $J$, and by testing also the possible new physics which could explain the negative value for $|U_{e3}|^2$ obtained in \cite{MSTV},  by using the presentday experimental data. For doing this we suggest the  use of  moduli, $a,\,b,\,c,\,d,\,e,\,f,\, g,\,h$ and   $J,\,\Delta_{12},\,\Delta_{13}$, as parameters,
although not all of them are truly independent. However the moduli are correlated through different forms for $J$ and/or $\cos\delta$, and relations as (\ref{uni}), or $J(a,b,c,e)=J(c,d,g,h)$ have to be satisfied to a great accuracy. And such a fit could provide values for all  moduli. If the new fit gives a positive  value for $J$ one can say that  the fit confirms the $CP$ violation in the leptonic sector. If the  fit will confirm the results from Ref.\,\cite{MSTV}, see its Appendix  A,  concerning the best fit obtained for negative values of  $|U_{e3}|^2,$ which imply that $a$ and $b$ entering the formula (\ref{real3}) for $P_{ee}$ are such that  $1-a^2-b^2 <0$, one can use the unitarity constraints to obtain other correlations. For example the   relation (\ref{real4}) implies also that $1-a^2-c^2=|U_{\tau 1}|^2 <0$, because otherwise  $e^2$ gets complex. If this is true then by   using  as independent parameters $a,\,c,\,e,{\rm and}\,f$, and eliminating again the parameter $e$, one  gets
\begin{widetext}\begin{eqnarray}
e^2=\frac{(a^2+c^2)(a^2-c^2 f^2)-(a^2-c^2)f^2}{(a^2+c^2)^2}+
\frac{2\epsilon\sqrt{a^2c^2f^2(1-a^2-c^2)(a^2+c^2-f^2)-(a^2+c^2)^2 J^2}}{(a^2+c^2)^2}\label{real5}
\end{eqnarray}\end{widetext}
Because $|U_{\tau 1}|^2= 1-a^2-c^2 <0$, the preceding relation tell us that $|U_{\tau 2}|^2=a^2+c^2-f^2 <0$ implying a new physics; in fact a new type of {\em unitarity}. In any case by using the proposed fit there is  a bigger flexibility to test various consequences of the theoretical model (\ref{mns}).

Of course results such  as $|U_{\tau i}|^2<0$, $i=1,\,2$, or  $|U_{e 3}|^2<0$ could be  artifacts of approximations used for the transition probabilities by neglecting the sub leading terms, and before speaking of new physics one must  use exact theoretical forms for all observables measured in neutrino physics. The unitarity is a very powerful tool because it implies many constraints between the moduli, and  even a moduli variation of the order of 2-3\%  could spoils it. See in this respect the numerical matrix (\ref{num}), and/or the matrix (\ref{w}), obtained from physical values for $s_{ij}$ and $\delta$, which show easily how one can walk into a trap.
Such a fit will provide values for all  moduli even if the nowadays data come only from three sources, and it deserves to be done since it will give estimates for all  transition probabilities necessary for the design of future neutrino factories.
 However, if such  a result is confirmed this will imply a new physics which could be interpreted as   the first sign  showing an  evidence for a new generation of leptons, or the change of the theoretical frame (\ref{mns}), by taking  the matrix $U$ as being   unitary, for example, in a non-Euclidean metric.

\section{Conclusions}

In this work we have discussed the question of the relevance of the nowadays experimental neutrino data on the determination of  $CP$ nonconservation in the leptonic sector, as well as on solving the negativity problem of $|U_{e3}|^2 <0$, raised by the best fit in Ref.\,\cite{MSTV}. For doing that we used another consequence of  unitarity property and defined a phenomenological model by taking as free parameters the moduli of the unitary matrix, which are invariant quantities, and which, naturally, leads to the separation criterion between the double stochastic matrices and those arising from unitary ones. This criterion provided  the necessary and sufficient conditions for the consistency of experimental data with the theoretical model encoded by the PMNS matrix,  the strongest condition being $-1\le\cos\delta \le1$. 

We constructed an exact double stochastic matrix, which from an experimental point of view is perfectly acceptable, but which does not come from a unitary matrix, showing the necessity of using the  separation criterion in phenomenological analyzes in order to obtain physically meaningful quantities. 

Taking into account the above result, we provided a reliable algorithm for the reconstruction of a unitary matrix from experimental data, when they are consistent with the theoretical model. The fitting algorithm has the form a least squares  $\chi^2$-test and have two separate pieces: one which implements the unitarity constraints, and the other which properly takes into account the error affected data. We have also shown how this algorithm can be modified to take into account the oscillation probabilities that are the primary data measured in neutrino physics.

We have used the  $\chi^2$-test
 to obtain information on the four parameters entering the standard parametrization of the PMNS matrix. The test  works very well, and by applying it we obtained a continuum of (approximate) unitary matrices compatible with the nowadays available experimental data summarized in the numerical matrix (\ref{mns1}). Their explicit construction shows that the moduli data (\ref{mns1}) are compatible with the existence of $CP$ violation in the leptonic sector, even  if one  cannot  yet find a precise value for phase $\delta$. 

 The  numerical results show that unitarity is a very strong property, and  the tuning of  moduli implied by it is given  at a  higher level than expected, the statistical errors generated by the approximate character of data are at least an order of magnitude  less than the experimental errors.

It was also shown that the fitting method,  using as independent parameters moduli and Jarlskog invariant, can directly check the existence of  $CP$ violation in the presentday neutrino data. On the other hand, the same method can resolve the positivity problem for  $|U_{e3}|^2$. If the results by Maltoni {\em et al}. \cite{MSTV} are confirmed, they will be the first signal for a new physics in leptonic sector.

Last but not least we used the natural embedding (\ref{ds2}) of unitary matrices into the double stochastic set for devising a method for doing statistics   on  unitary matrices, problem that was an open one until now. The method allowed us to find a procedure to obtain mean values, and,  correspondingly, error matrices starting with a set of (approximate) unitary ones. 

As a final conclusion, a satisfactory solution of $CP$ nonconservation can be obtained by diminishing the errors in all running experiments, as well as in all future ones, the present theoretical approach being able to test the validity of the nowadays unitary model, or the necessity to modify it.

\begin{acknowledgments} 
 I thank also L Lavoura and C Hamzaoui for pointing out their papers to me.\end{acknowledgments}

\end{document}